# Roman CCS White Paper
# RoSETZ: Roman Survey of the Earth Transit Zone – a SETI-optimized survey for habitable-zone exoplanets

Eamonn Kerins[1,*], Supachai Awiphan[2], Kathryn Edmondson[1], Michael Garrett[1], Jacob Haqq-Misra[3], Réne Heller[4], Macy Huston[5,6,7], David Kipping[8], Ravi Kopparapu[9], Danny C. Price[10], Andrew Siemion[1,11,12], Siddhant Sharma[3,13], Evan L. Sneed[7,11,14], Hector Socas-Navarro[15,16], Robert F. Wilson[9], and Jason Wright[5,6,7]

[1]Jodrell Bank Centre for Astrophysics, Department of Physics and Astronomy, The University of Manchester, Oxford Road, Manchester M13 9PL, UK.
[2]National Astronomical Research Institute of Thailand, 260 Moo 4, Donkaew, Mae Rim, Chiang Mai, 50180, Thailand
[3]Blue Marble Space Institute of Science, Seattle, 98104 WA, USA
[4]Max Planck Institute for Solar System Research, Justus-von-Liebig-Weg 3, 37077 Göttingen, Germany
[5]Department of Astronomy and Astrophysics, 525 Davey Laboratory, The Pennsylvania State University, University Park, PA, 16802, USA
[6]Center for Exoplanets and Habitable Worlds, 525 Davey Laboratory, The Pennsylvania State University, University Park, PA, 16802, USA
[7]Penn State Extraterrestrial Intelligence Center, 525 Davey Laboratory, The Pennsylvania State University, University Park, PA, 16802, USA
[8]Department of Astronomy, Columbia University, New York, NY, 10027, USA
[9]NASA Goddard Space Flight Center, 8800 Greenbelt Road, Greenbelt, MD 20771, USA
[10]International Centre for Radio Astronomy Research, 1 Turner Ave, Bentley WA 6102, Australia
[11]Breakthrough Listen, University of California, Berkeley, CA 94720, USA
[12]SETI Institute, Mountain View, CA 94043, USA
[13]School of Chemistry, University of New South Wales, Sydney, NSW 2052, Australia
[14]Department of Earth and Planetary Sciences, University of California, Riverside, CA 92521, USA
[15]Instituto de Astrofísica de Canarias, Vía Láctea S/N, La Laguna 38205, Tenerife, Spain
[16]Departamento de Astrofísica, Universidad de La Laguna, 38205, Tenerife, Spain
*Corresponding author: Eamonn.Kerins@manchester.ac.uk

## ABSTRACT

In this White Paper for Nancy Grace Roman Space Telescope (Roman) science, we propose the *Roman Survey of the Earth Transit Zone* (RoSETZ), a transit search for rocky planets within the habitable zones (HZs) of stars located within the Earth Transit Zone (ETZ). The ETZ holds special interest in the search for extra-terrestrial intelligence (SETI) - observers on planets within the ETZ can see Earth as a transiting planet. RoSETZ would augment the Roman Galactic Bulge Time Domain Survey (GBTDS) as an additional field located ∼ 5 degrees away from other GBTDS fields. Our simulations show that RoSETZ alone can find from 120 to 630 Earth-sized HZ planets around K- and M-type hosts, with the range reflecting different survey design assumptions. These yields are 5-20 times the number currently known. Such a sample will transform our knowledge of "Eta-Earth" ($\eta_\oplus$) – the occurrence of Earth-sized HZ planets – and would be the first catalogue of exoplanets selected in a manner optimized according to the Mutual Detectability targetted-SETI strategy. If it can be accommodated alongside the existing GBTDS design, we favour a RoSETZ-Max design that is observed for the duration of the GBTDS. If not, we show that a slimmed-down RoSETZ-Lite design, occupying two GBTDS seasons, would not significantly impact overall GBTDS exoplanet yields, even if time allocated to it had to come from time allocations to other fields. We argue that the angular separation of RoSETZ from other GBTDS fields permits self-calibration of systematic uncertainties that would otherwise hamper exoplanet demographic modelling of both microlensing and transit datasets. Other science possible with RoSETZ data include studies of small solar system bodies and high resolution 3D extinction mapping.

**SCIENTIFIC CATEGORIES:** Exoplanets and exoplanet formation; Stellar populations and the interstellar medium

**KEYWORDS: Exoplanets And Exoplanet Formation:** Exoplanet detection methods; Transits; Exoplanet systems; Extrasolar rocky planets. **Stellar Physics and Stellar Types:** Low mass stars. **Search for Extra-terrestrial Intelligence:** targeted SETI.

## 1 The Roman Road to SETI-optimal exoplanet science: minimizing risk, maximizing gain

The Vision of NASA's Science Mission Directorate lists six questions that the directorate seeks to address[1]. Two of these are: "*How do planets and life originate?*"; and "*Are we alone?*". These are fundamental questions that seek to understand just how remarkable it is that Earth is a world capable of sustaining life, and that life, indeed intelligent life, has emerged on it. Is our existence all but inevitable given the right conditions? Or are we the result of a most incredible sequence of chance events that is not likely to be repeated elsewhere? Currently, we have no firm answers, but with the Nancy Grace Roman Space Telescope (hereafter Roman, Spergel et al. 2015) we have a great opportunity to take a big leap forward in our understanding.

The field of exoplanets is a remarkable success story. In less than thirty years, an array of methods has been used to find almost 5,500 confirmed exoplanets[2]. The NASA Kepler mission (Borucki et al., 2010) has played a huge role, with its detection of transiting planets accounting for more than half of all confirmed exoplanets. The Transiting Exoplanet Survey Satellite (TESS, Ricker et al. 2015) is currently the most prolific transit search mission, which so far has found 346 confirmed planets, and over 6,000 additional candidates awaiting verification. The upcoming ESA PLATO mission will also use the transit method and be able to search for planets located within the habitable zone (HZ) of their host (Rauer et al., 2014). Missions like Kepler, TESS and PLATO are testament, not just to the power of the transit detection method for exoplanet discovery and characterization, but also to the huge scientific advantages of conducting such surveys from space.

Analyses of transit and radial velocity surveys show that the occurrence of exoplanets around stars is high, especially for small planets around low-mass stars (e.g. Hardegree-Ullman et al., 2019; Bryson et al., 2021; Sabotta et al., 2021; Pinamonti et al., 2022). Indeed, we now know that essentially all stars host planets and that Earth-sized planets are among the most common. Surveys are also showing that there is a relatively high occurrence of planets within the HZ of their host star (Bryson et al., 2021). The HZ perhaps represents a somewhat conservative view of the region around stellar hosts within which life could emerge and be sustained. Even in our own Solar System, we are learning that it may be possible for life to be hosted on bodies, like Europa, that reside well outside the HZ. And so the habitability of moons beyond the solar system, so-called exomoons, seems plausible (Heller et al., 2014).

These discoveries have boosted interest in the search for life, including the search for extra-terrestrial intelligence (SETI). In this White Paper we use the term SETI as a synonym for the hunt for technosignatures. A further boost to SETI research has come from Breakthrough Listen (Price et al., 2020; Sheikh et al., 2020; Garrett & Siemion, 2023), a huge philanthropically-funded program to look for evidence of intelligent civilizations beyond Earth. Breakthrough Listen is using multiple radio and optical observatories, coupled with modern data mining approaches, for the task of looking for evidence of technosignatures. Whilst SETI research has been traditionally regarded as a "high risk" endeavour, Breakthrough Listen has helped to foster a better appreciation among the astrophysics community for the discovery potential of technosignature surveys for many areas of astrophysics (Lacki et al., 2021), both because of the novel observational approaches that such surveys use, and because of new data analysis techniques that are developed for them.

The Roman mission will undertake ground-breaking surveys across a wide gamut of astrophysics through its three Core Community Surveys, its General Astrophysics program and its coronagraph demonstrator program.

---

[1] https://science.nasa.gov/about-us/smd-vision/
[2] As of June 2023: https://exoplanetarchive.ipac.caltech.edu/

The Galactic Bulge Time Domain Survey (hereafter GBTDS) is a Roman Core Community Survey devoted to studying planetary and stellar populations towards the inner Galaxy. GBTDS will be a sustained, high temporal cadence, near-IR photometric survey of a $\sim 2$ deg$^2$ region close to the Galactic Centre. The principal objective of the GBTDS is to conduct a survey for cool, low-mass exoplanets using the microlensing detection method (Penny et al., 2019). The GBTDS will probe an exoplanet demographic that was inaccessible to Kepler, and indeed inaccessible to detection techniques other than microlensing. The planetary regime is of central importance to our understanding of planet formation and its connection to present-day exoplanet architectures. Cool, low-mass exoplanets are predicted not to migrate after formation (Burn et al., 2021); to access this regime is to access an exoplanet demographic that provides a fossilized signature of the planet formation process.

In order to achieve its goals, the GBTDS will survey one of the densest regions of the Galaxy with a nominal seven pointings of the Roman Wide Field Imager (WFI). Observations of each field will be repeated every 15 minutes for continuous seasons of 60-72 days duration. Six monitoring seasons, totalling 360-432 days, are envisaged over the 5-year nominal Roman mission. As well as the discovery of potentially $\sim 1,400$ cool exoplanets, the GBTDS survey design will result in the detection of 60,000-200,000 distant transiting planets (Wilson et al., 2023). This catalogue will be an unrivalled gold mine for demographic studies and will enable the microlensing and transit samples to be combined to facilitate studies of how planetary architecture varies with Galactic location.

The purpose of this White Paper is to consider the science potential of a Roman survey field that is separated by around 5 degrees from the nominal GBTDS survey area. Its location would fall within the Earth Transit Zone (ETZ; Heller & Pudritz 2016), a region straddling the Ecliptic plane within which observers on other planets could observe Earth as a transiting planet. We refer to this offset field as the Roman Survey of the Earth Transit Zone (RoSETZ). Apart from its offset location, RoSETZ would otherwise share the same basic design as the rest of the GBTDS. We argue that the wide field of view of Roman is near optimal to provide a very efficient survey for exoplanets located within the ETZ.

RoSETZ would be designed to deliver a sample of Earth-sized, HZ transiting planets orbiting around the most common star types. But, uniquely for an exoplanet survey, this sample would be optimized for future SETI follow-up programs along principles put forward in a recently proposed, game-theory motivated SETI strategy (Kerins, 2021). Through RoSETZ, Roman will simultaneously deliver ground-breaking exoplanet science, whilst enabling new smart approaches to technosignature surveys. We anticipate that the planets found by RoSETZ will spark significant public interest, as life beyond Earth is widely recognised as an area of science that inspires huge public interest and engagement.

We propose an ambitious version of RoSETZ (RoSETZ-Max) that could be added to the existing GBTDS if Roman slew and settle times are sufficiently fast to allow its inclusion without impacting the GBTDS exoplanet science goals. Otherwise, we show that a more modest proposal, RoSETZ-Lite, can be delivered with very minimal impact to the microlensing survey. Indeed, we show that having a separated field is beneficial in providing self-calibration of systematic model uncertainties in the analyses of both GBTDS microlensing and transit samples. We use simulations to show that RoSETZ alone would deliver a yield of rocky HZ planets around K- and M-type hosts that exceeds the current sample by a factor of between 5 (RoSETZ-Lite designs) and 20 (RoSETZ-Max designs). We also briefly outline examples of other science opportunities that the RoSETZ field would present, including the construction of deep, detailed 3D extinction maps and searches for small solar system bodies.

The White Paper is set out as follows. In Section 2 we outline the idea of Mutual Detectability as a smart strategy for SETI, motivated by game theory considerations. We also discuss the importance of the ETZ for this strategy. In Section 3 we define the parameters and variations of the RoSETZ survey design. In Section 4 we present detailed transit simulations to compute the expected yield of rocky HZ planets around low-mass stellar hosts. We explore in Section 5 how RoSETZ can provide self-calibration of systematic model uncertainties for the GBTDS microlensing and transit exoplanet samples. Some additional science possibilities that would be enabled by RoSETZ are presented in Section 6.

## 2 SETI as a two-player game: Mutual Detectability and the Earth Transit Zone

In his 1960 book, *The Strategy of Conflict*, Cold War strategist Thomas Schelling outlined a novel approach to the strategic management of conflict, bargaining and cooperation. His pioneering approach helped to spark the field of game theory and would ultimately contribute to him being awarded the Nobel Prize for Economics in 2005. Among the many situations first proposed in Schelling's book is the problem of the *Strangers Meeting in New York*. This problem poses the question of what is the best strategy to take for two people, who have never met and cannot communicate with each other, to meet up on a particular day somewhere in New York City. They don't know where in the city to meet, or at what time. It is perhaps the most famous game theory problem involving a cooperative approach by two non-communicating participants whose only knowledge about the other is that they share a mutual objective. As Schelling himself acknowledges in a footnote in the book, strategies for two-player non-communicative games of cooperation are relevant to the endeavour of SETI.

Schelling's innovative solution to the problem of the Strangers Meeting in New York demonstrates that, what initially may seem impossible odds of success, can be overcome (or, at least, can become rather more possible) if both parties take a game theory approach to the problem and attempt to think about what the other person might do. Schelling advocates that the most successful approach would be to select a meeting place where people often arrange to meet up (eg the lobby of Grand Central station), and a meeting time that people often arrange to meet at (e.g. Noon). If both parties adopt such a strategy, success may still remain elusive, as there are many possible combinations of such choices, but a successful outcome becomes far from impossible. What this highlights is that our decisions in such situations are not random; there are choices we make that we favour over other possibilities. Such choices have come to be known in game theory as "Schelling points". The identification of suitable Schelling points by both participants is key to maximizing the chance of a successful outcome. For SETI, our best chance of contact with another civilization may come if we both recognize that we are playing a non-communicative, two-player game of cooperation towards the mutual goal of contact. To maximize the chance of mutual success, we must both identify suitable Schelling points for where and when to look.

As highlighted by Wright (2018), many of the strategies adopted by technosignature surveys can be considered Schelling points. For example, a historically common choice for SETI surveys has been to look for narrowband radio transmissions near the 21 cm emission wavelength of neutral hydrogen. The emission wavelength of hydrogen is a fact presumably known to any civilization of comparable or superior technology to our own, and therefore would be among a list of Schelling-point transmission wavelengths for them to consider, if they wished to maximize their chance of being noticed. As a more recent example of a SETI Schelling point, Kipping & Teachey (2016) point out that a civilization could decide to broadcast their presence to others by emitting a laser signal towards a planet located close to their orbital plane at an epoch where the emitter's planet begins to transit its host star, an epoch at which observers on the other planet may be most likely to look their way. This would constitute both a temporal (when?) and spatial (where?) Schelling point. Heller & Pudritz (2016) considered the problem from the opposite perspective by estimating how many stars are located within Earth's Transit Zone (ETZ). The ETZ is the thin band straddling the Ecliptic plane from which outside observers can see Earth's annual transit of the Sun. They concluded that there should be $\sim 10^5$ K- and G-type dwarfs within the ETZ. Civilizations located around any of these systems could be motivated by their observation of Earth's transit signal to transmit a signal to us.

Kerins (2021) proposed the idea of Mutual Detectability as a game-theory based smart-strategy for designing targeted SETI searches. The idea of Mutual Detectability is that a pair of civilizations will have a greater chance of establishing contact when they are able to recognize mutually available basic evidence of each other's potential existence. This evidence, and the fact that both parties recognize that is it mutually shared, may provide sufficient incentive for at least one party to transmit a signal to the other. Whilst the idea of Mutual Detectability represents a general framework, Kerins (2021) applied these principles to the situation of transiting planets that are located within the ETZ. In this case, observers at both ends can know that the other can view their home planet transit their host star. Transits are intrinsically simple planet detection signals that can be accessible even to civilizations like ourselves that are only recently capable of finding exoplanets via any method. But yet transit signals also provide a lot of information about the potential habitability of a planet. Transit data can establish the size and temperature of the planet and even establish the chemical composition of the atmosphere. Whilst we are not quite at the point of being able to do so, observers with technology a few decades ahead of our own may be able to use the transit of Earth to determine the presence of biomarker signatures in Earth's atmosphere or even, if they are sufficiently close, evidence of technosignature pollutants (Haqq-Misra et al., 2022).

This remarkable combination of intrinsic simplicity, coupled with an ability to convey a lot of information about habitability, make transit signals ideal Schelling points. The greater the scrutiny of Earth by observers located within the ETZ, the more information they gain of the existence of life here, and potentially even of intelligent life. If their planet is also visible to us as a transiting planet, they will also be aware that we could gain similar access about their potential existence. In a case where both parties desire contact, the very recognition that such information is mutually available forms a strong incentive for one, or both, to consider transmitting to the other.

The most basic quantitative piece of information that both parties can know, without the need for prior communication, is the overall signal strength of each other's transit signal. Kerins (2021) shows that the transit signal strength of a planet we observe, relative to Earth's transit signal, is

$$\frac{S}{S_\oplus} = \frac{L_*}{L_\odot} \left(\frac{R_*}{R_\odot}\right)^{-2} \left(\frac{R_\mathrm{p}}{R_\oplus}\right)^2 \left(\frac{P}{\mathrm{yr}}\right)^{-1} \frac{t_{14}}{t_\oplus}, \qquad (1)$$

where $t_\oplus = 12.9$ hours is the maximal duration of Earth's transit, $L_*$ and $R_*$ are the luminsoity and radius of the star we see being transited, $R_\mathrm{p}$ and $P$ are the radius and orbital period of the transiting planet, and $t_{14}$ is its transit duration. The Earth's transit signal strength is $S_\oplus = 10^3$ $L_\odot$ ppm hours yr$^{-1}$. The ratio $S/S_\oplus$ is intrinsic and so independent of the technological capabilities of either side. Both $S$ and $S_\oplus$, and therefore their ratio, are mutual information, that is to say knowable to both sets of observers. When $S < S_\oplus$ both parties (them and us) know that Earth's intrinsic signal is stronger. Kerins (2021) argues that, in the event that both sides desire contact, observers looking at Earth may be incentivized to initiate contact in a situation where both they and us know that the signal they see of us is stronger. Using piecewise fits for the stellar main sequence mass-radius-luminosity relation from Eker et al. (2018), the requirement that this be true is ensured by a remarkably simple criterion:

$$\frac{R_\mathrm{p}}{R_\oplus} \lesssim \left(\frac{L_*}{L_\odot}\right)^{-1/7}. \qquad (2)$$

Therefore, under the strategy of Mutual Detectability, observers living on rocky (i.e. roughly Earth-sized) planets located within the ETZ have greater incentive to transmit to us, than vice-versa, if their planet orbits around a sub-solar luminosity host.

The recommendation of this approach is then for us to assemble a catalogue of rocky HZ transiting planets of sub-solar luminosity hosts that are located within the ETZ. Future targeted SETI surveys would be able to conduct technosignature searches towards these systems, such as looking for evidence of laser emission during transits (Kipping & Teachey, 2016).

The assembly of a SETI-optimized catalogue of rocky HZ planets is one of the central aims of the RoSETZ proposal, combining ground-breaking exoplanet discovery with the enabling of new smart strategies for future SETI programs. RoSETZ will facilitate high precision measurements of the occurrence of Earth-sized HZ planets orbiting the most common stars, whilst opening up an exciting new avenue for future tecnosignature investigations.

## 3 Pinning a RoSETZ onto the Galactic Bulge Time Domain Survey

RoSETZ would use the same broad $0.9 - 2$ $\mu$m WFI F146 filter that will be the principal filter for GBTDS observations. As we shall discuss in Sections 5.1 and 5.2, the RoSETZ field would itself provide transit and exoplanetary microlensing samples that can provide self-calibration of systematic uncertainties in exoplanet demographic modeling of GBTDS datasets.

The Roman WFI provides a number of unique capabilities that RoSETZ would exploit for finding large numbers of Earth-like HZ planets located in the ETZ. Firstly, as illustrated in Figure 1, the large 0.281 deg$^2$ WFI active aerial coverage[3] is highly compatible with the $0°.53$ width of the ETZ (Heller & Pudritz, 2016). Even in a worst-case scenario, where the WFI is aligned with its long axis traversing the ETZ, a fraction $f_{\text{ETZ}} \geq 0.66$ of the planets within the WFI field will also lie within the ETZ. In reality, there is some leeway to rotate the WFI within the limits set by the solar aspect angle constraint, so $f_{\text{ETZ}} = 0.66$ is a pessimistic lower bound. In this white paper we normalize all of our simulation yields to $f_{\text{ETZ}} = 1$.

Secondly, the Roman broad F146 filter has excellent sensitivity into the near-IR, enabling it to be effective even in regions of substantial optical extinction. The region of the ETZ that intersects with the Galactic plane is a high extinction area (see Figure 1), yet our simulations show that Roman will be able to detect many hundreds of HZ Earth-sized planets out to $\sim 4$ kpc. Indeed, as we discuss, Roman data will enable the development of deep 3D maps of interstellar extinction that can surpass the fidelity of those from previous Galactic bulge surveys.

Thirdly, Roman's exquisite spatial resolution will provide superlative photometric and astrometric performance that is critical for accurate transit measurements of Earth-sized planets. The transit yield from the ETZ field will be comparable to the fields within the nominal GBTDS area, despite being a factor 2 lower in stellar number density. This is because the GBTDS field suffer more from blending effects, which affects both yield and also the accuracy of transit depth (and therefore planet radius and mass) recovery.

### 3.1 RoSETZ Survey Designs: RoSETZ-Lite and RoSETZ-Max

Here, we set out two example survey designs for RoSETZ. **RoSETZ-Lite** is a single-field, two-season survey of a field located within the ETZ, with the two seasons separated by 4.5 years to facilitate transit deblending and microlensing proper motion measurements. **RoSETZ-Max** is a single-field, six-season survey of the ETZ, surveyed whenever the GBTDS is surveyed. Our example field is centred on the intersection of the Galactic and Ecliptic planes at $l = 6°.4$, $b = 0°$. A thorough evaluation of the optimal location would be needed to maximize the yield; other nearby locations within the ETZ with lower extinction may well provide more detections, so our forecasts here should be viewed as conservative.

We envisage RoSETZ-Max to be the design of choice if it can be accommodated alongside the GBTDS without compromising its science goals. There is an expectation that the Roman slew and settle time will turn out to be rather faster than originally anticipated, in which case it may be possible to add another field whilst maintaining survey requirements for the nominal GBTDS campaign. In this case RoSETZ-Max would add an eighth field, albeit separated from the other seven, with otherwise the same monitoring strategy.

In the case that the GBTDS would be impacted at some level by the addition of an extra field, we propose that RoSETZ-Lite may offer a good compromise. We show in Section 5.1 that a two-season campaign towards the ETZ should not drastically alter the overall achievable microlensing yield, even if it takes time away from the nominal GBTDS region. We argue that having a well-separated field is, in fact, desirable in order to control exoplanet demographic systematics.

---

[3]Here and elsewhere, unless stated otherwise, Roman telescope, detector and filter characteristics are obtained from https://roman.ipac.caltech.edu/sims/Param_db.html?csvfile=RomanParameters_Phase_C_01-20-2023.csv, accessed May/June 2023.

[4]https://aladin.cds.unistra.fr/

[5]Reading from the plots contained in https://roman.gsfc.nasa.gov/science/2020-01/wim_psf_subset.zip, accessed May/June 2023.

[6]Values taken from the tables in https://roman.gsfc.nasa.gov/science/WFI_technical.html, accessed in May/June 2023.

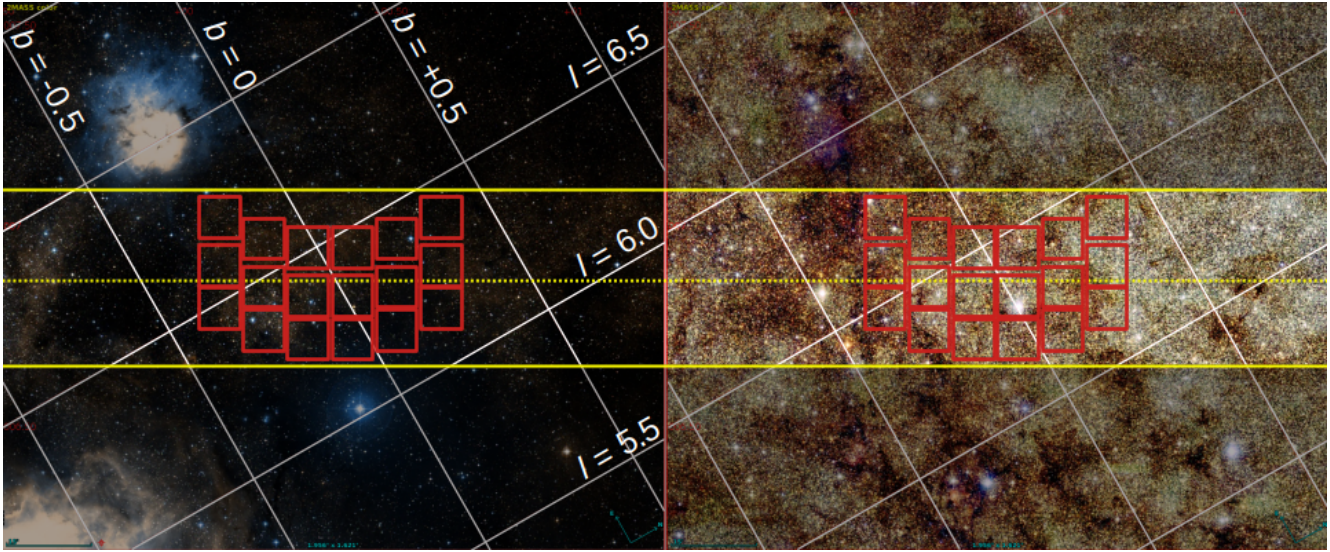

**Figure 1.** An Aladin Sky Atlas[4] view of the ETZ centred at $l = 6°.4$, $b = 0°$, where the Ecliptic and Galactic planes intersect. Left panel shows a DSS optical view whilst the right panel provides a near-IR JHK 2MASS view. The optical view is obscured by extinction and dominated locally by Messier 20 (the Trifid Nebulae), which is visible towards the top left. Roman's view will resemble that of the JHK 2MASS image, though will go much deeper and with much higher spatial resolution. The sky views are both orientated to the Ecliptic plane with the yellow dotted line showing the Ecliptic and the solid yellow lines showing the boundaries of the ETZ. Observers on planets located within this region can view Earth's annual transit of the Sun. White lines show the Galactic coordinate grid. The Roman WFI detector footprint is shown to approximate scale aligned along the Ecliptic to illustrate its aerial compatibility with the span of the ETZ. In practise its orientation will be restricted by the operational limits imposed by the solar aspect angle though, even if orientated perpendicular to the ETZ, at least 66% of the detector area will lie within the ETZ.

| RoSETZ parameters - all simulations | |
|---|---|
| Field centre ($l, b$), J2000 | (6°.4, 0°) |
| Assumed fractional overlap of WFI area with ETZ, $f_{ETZ}$ | $f_{ETZ} = 1$ |
| Primary filter | WFI F146 |
| Exposure time | 55 secs |
| Observing cadence | 15 mins |
| F146 mag zeropoint | 27.648 |
| F146 background count rate | 1.83 sec$^{-1}$ |
| Photometry | 2-pixel radius aperture photometry, corresponding to 70% encirled energy of the F146 point spread function[5] |
| Photometric noise model | Poisson, with a systematic 1 mmag precision floor added in quadrature |
| **RoSETZ survey design variations** | |
| Number of observing seasons | RoSETZ-Lite: 2, with a 4.5-year separation<br>RoSETZ-Max: 6, with 2 yr$^{-1}$ in years 1, 2 and 5 |
| Observing season duration | 60 or 72 days |

**Table 1.** Parameters used for all of our RoSETZ simulations, and also survey design variations considered. The adopted values for the exposure time, observing cadence, F146 zero-point magnitude, and photometric precision floor are based on those in Penny et al. (2019) and Wilson et al. (2023). The background count rate includes detector thermal noise and a zodiacal light contribution set at 5x minimum value[6]. The background from unresolved blended stars is computed and accounted for within our simulation. The impact of both short (60-day) and long (72-day) observing seasons is evaluated for both RoSETZ-Lite and RoSETZ-Max.

| Transit simulation distributions and parameters | |
|---|---|
| Galactic model | Besançon Galactic population synthesis model V1612[7] |
| Semi-major axis distribution, $f(a)$ | $df(a)/d\ln a =$ constant, circular orbits |
| Planet-host mass ratio distribution, $f(q)$ | $df(q)/d\ln q \propto (q/q_{br})^x$  $(3 \times 10^{-6} < q < 8 \times 10^{-5})$, $x = 1$  $(q < q_{br})$,  $x = -2.8$  $(q \geq q_{br})$, $q_{br} = 2.8 \times 10^{-5}$ |
| Planet mass–radius ($M_p - R_p$) relation | $R_p/R_\oplus = (M_p/M_\oplus)^{0.28}$ |
| **Transit selection cuts** | |
| Host star types | Restricted to main sequence stars later than G5 |
| Planet–host separation | Required to be within the optimistic HZ: "Recent Venus" to "Early Mars" |
| Planet mass, $M_p$ | $0.1 \leq M_p/M_\oplus \leq 5$ |
| Planet period, $P$ | No larger than half the time between initial and final survey epochs: |
| Required min. number of transits, $N_{trans}$ | $N_{trans} \geq 2$, with transits required to be fully contained within observing seasons |
| Transit detection threshold | Apparent depth has signal-to-noise ratio, $S/N_{trans} \geq 8$ |

**Table 2.** Transit simulation distributions, parameters and selection cuts. The parameters used for $f(q)$ are based on those in Table 1 of Pascucci et al. (2018) for M and K-type stars. The extent of the optimistic HZ varies with host effective temperature following Equation (4) of Kopparapu et al. (2014). The planet mass–radius relation is applicable to rocky planets (Edmondson, Norris & Kerins, in prep). The $S/N_{trans}$ threshold value matches that used by Wilson et al. (2023). All simulations are normalized to $f_{ETZ} = 1$ and to an occurrence of one planet per star within the host's optimistic HZ.

The parameters adopted for all of our simulation work are summarized in Table 1.

# 4 Forecasts for RoSETZ

We have run detailed simulations to forecast the number of rocky HZ transiting planets detectable to RoSETZ. Assumed parameters for the Roman WFI detector are summarized in Table 1 and correspond closely to those used in extensive image-level simulations of the microlensing (Penny et al., 2019) and transit (Wilson et al., 2023) yields for the GBTDS. In this White Paper we have not performed image-level simulations, but we have included many of the technical factors that affect detection sensitivity, including detector thermal noise, zodiacal background light, starlight blended within the host point spread function (PSF), and cut-off of the PSF within an assumed 2-pixel photometric aperture. Our calculations are based conservatively on aperture photometry, whereas in practise the extremely well calibrated Roman WFI PSF should allow higher precision measurements using PSF fitting and/or difference image photometry.

To model the population of host and blend stars we have used the Besançon Galactic Model (BGM: Robin et al., 2003, 2012), incorporating a 3D extinction model (Marshall et al., 2006), to generate artificial catalogues of stars. We use the online BGM tool[7] to generate catalogues for the ETZ location at Galactic coordinates $l = 6°.4$, $b = 0°$. The BGM is a state-of-the-art synthetic population synthesis model of the Galaxy that has been developed over many decades and is under continuous revision. Since its initial use for microlensing predictions (Kerins et al., 2009), it has been used extensively to forecast space-based microlensing and transit yields (Penny et al., 2013; McDonald et al., 2014; Penny et al., 2019; Wilson et al., 2023). The artificial stellar catalogues list a wide range of stellar properties, including: magnitudes in a range of filter systems; mass, radius and effective temperature; proper motion; distance and foreground extinction. We transform from *JHK* magnitudes to the wide Roman WFI F146 filter using the recipe in Wilson et al. (2023).

---
[7]Simulation input catalogues obtained from https://model.obs-besancon.fr/, accessed May/June 2023.

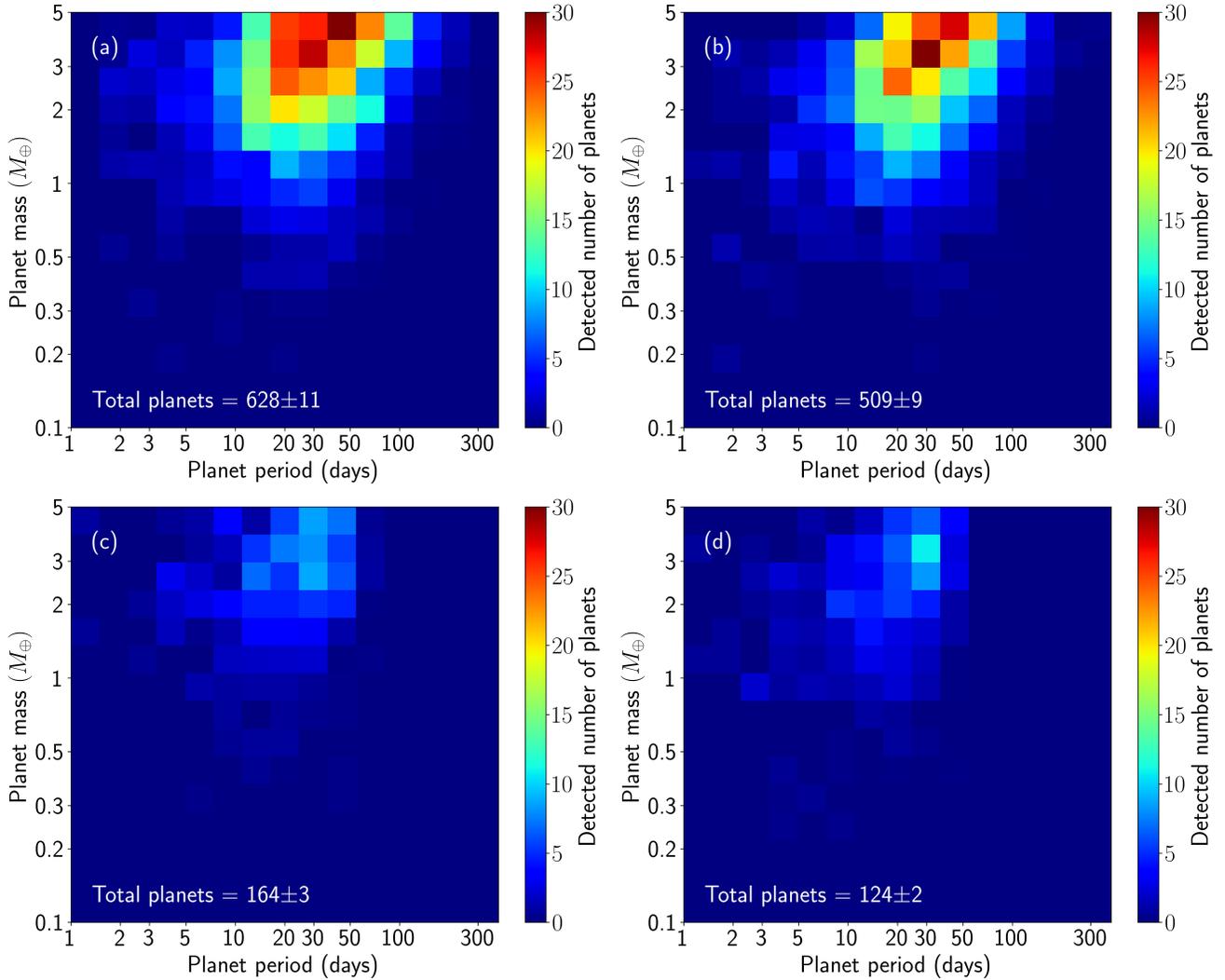

**Figure 2.** The number of simulated RoSETZ transit detections in the logarithmically-stretched plane of planet mass versus orbital period for a single Roman WFI field centred on the ETZ at $l = 6°\!.4$, $b = 0°$. Here, and elsewhere, the histograms are normalised to $f_{\mathsf{ETZ}} = 1$ and assume that every star hosts one planet within its optimistic HZ. The projections are shown for the cases of (a) RoSETZ-Max with 72-day seasons, (b) RoSETZ-Max with 60-day seasons, (c) RoSETZ-Lite with 72-day seasons, and (d) RoSETZ-Lite with 60-day seasons. The histograms are determined from the average of 100 simulation runs, though using the same Besançon Galaxy Model input catalogues. The total yields compare favourably to the current sample of 32 known exoplanets, with similar mass and host insolation, that orbit K- or M-type hosts[8]. At least 66% of the RoSETZ sample will be located within the ETZ.

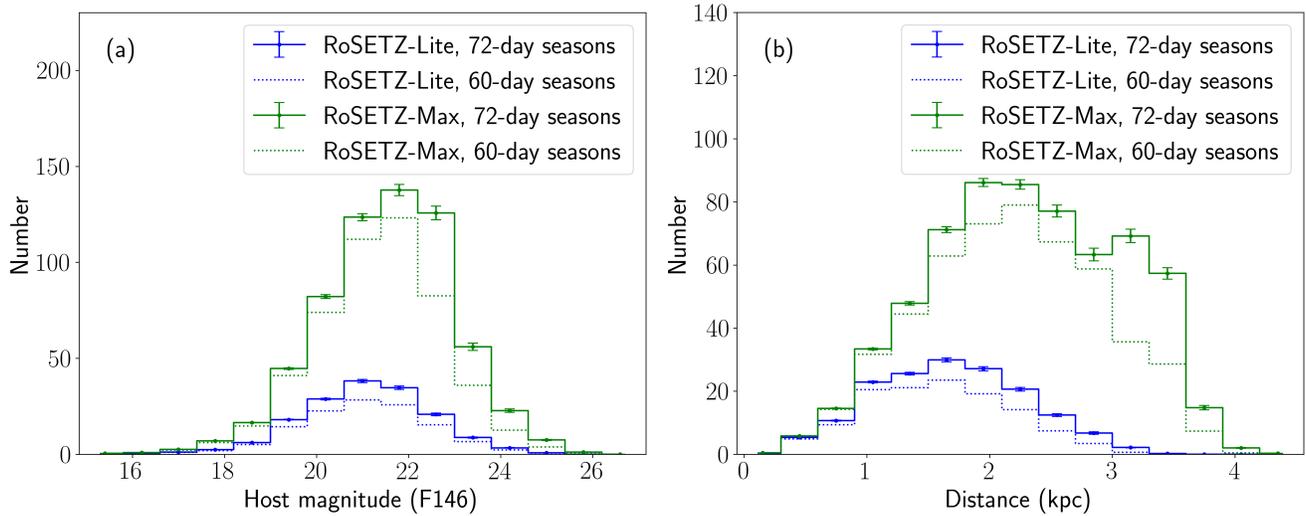

**Figure 3.** The number of simulated transit detections of HZ planets as a function of (a) WFI F146 magnitude, and (b) distance. The solid histograms show projections for RoSETZ-Lite (blue) and RoSETZ-Max (green), with each season lasting for 72 days. The error bars are derived from the statistical variance measured across 100 simulations. The dotted line histograms show corresponding projections for shorter seasons lasting 60 days, and with error bars omitted for clarity.

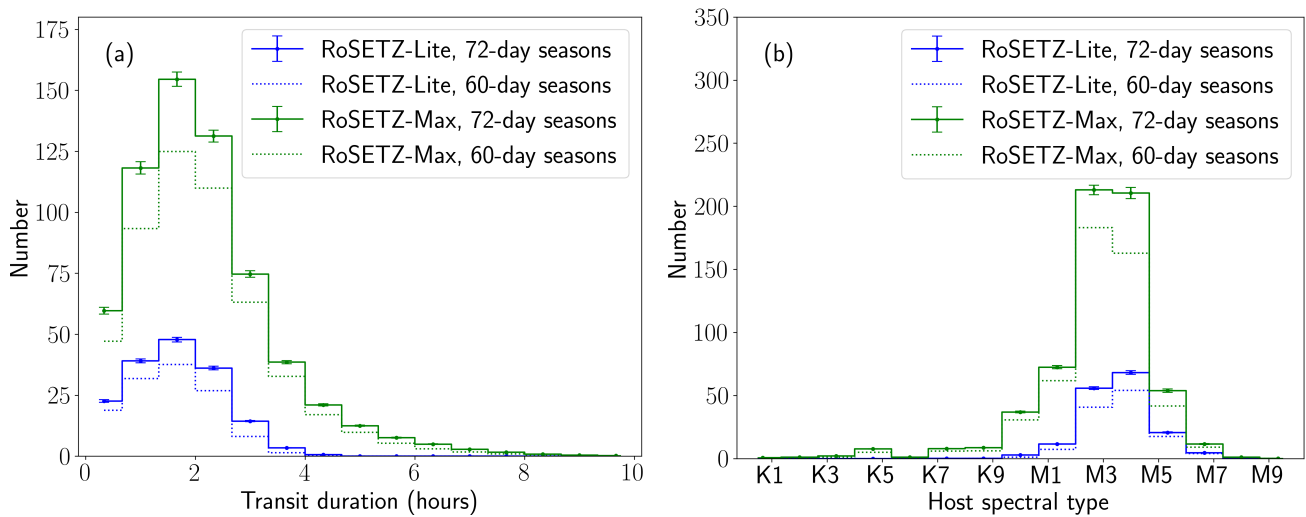

**Figure 4.** As for Figure 3, but showing (a) the transit duration, and (b) host spectral type of simulated detections.

For each input star, we generate a planet with characteristics drawn from the distributions given in Table 2, based on recent determinations of exoplanet demography (e.g. Pascucci et al., 2018). Appropriate weightings are computed based upon transit probability. We make a number of selection cuts to decide upon detectable transits; these are listed in Table 2. In particular, we are interested here in the transit yield of rocky HZ planets orbiting low-mass stellar hosts. Accordingly, our cuts include a restriction to hosts of type later than G5, planet masses below 5 $M_\oplus$, and planets orbiting within the optimistic HZ (Kopparapu et al., 2014).

We consider variations in season length for both RoSETZ-Lite and RoSETZ-Max spanning the durations that have been investigated to date by the Roman GBTDS science team. We deem a transit detection to occur when the "measured" transit depth (including dilution by stars blended into the host PSF) is at least 8 times larger than the quadrature combined error in the baseline and transit levels. Photometric errors on individual data points are determined from a Poisson noise model combined in quadrature with a systematic photometry floor of 1 mmag. We assume that the determination of baseline and transit levels benefit from a $1/\sqrt{N}$ improvement in precision due to the number, $N$, of data points that populate the transit, and the baseline, respectively. We ignore the effects of host limb darkening on the transit profile, though these are expected to be weak for the F146 near-IR filter. The overall simulation projection, and its error, are determined from the mean and variance of 100 simulation runs for each RoSETZ survey variation. We also perform a simulation centred at $l = 0°\!.5, b = -1°\!.5$, corresponding to a representative location within the GBTDS, in order to compare and contrast results from the RoSETZ field to those from the nominal GBTDS area.

Figure 2 shows the distribution of simulated detections in the plane of (true) planet mass versus (true) period. The overall detection yield for all of our survey variations ranges from around 120 (two-season RoSETZ-Lite with 60-day seasons) to around 630 (6-season RoSETZ-Max with 72-day seasons). The vast majority involve M-type hosts, with a small number of K-type hosts. Although G-type hosts later than G5 were not excluded from the simulation, there were usually no simulated detections of HZ transiting planets around them. This is both because G-type stars are intrinsically rare compared to M- and K-types and because the longer orbital periods provide fewer transit signals that are also shallower for these larger stars. The RoSETZ yield forecasts should be compared against the current total of 32 confirmed exoplanets below 5 $M_\oplus$ orbiting within the optimistic HZ of K- or M-type hosts.[8] Both RoSETZ, and indeed the GBTDS, will clearly be revolutionary for the determination of $\eta_\oplus$, a statistic that will be dominated by K- and M-type hosts as they are the most common stars in the Galaxy.

Our simulations find that, for short observing campaigns (one or two seasons), the difference between a 60 and 72-day continuous observing window can have a disproportionately large effect on yield, whilst for large multi-season campaigns the numbers scale more proportionately. For example, we find there is a factor three difference in overall yield between RoSETZ-Lite and a survey that would run for only a single season. This is because there are many planets with periods above the half-season duration that are too long to register at least two transits within a single season. This, together with the additional astrometric information that a second, well-separated, observing season would bring, motivates our requirement that RoSETZ-Lite should span a minimum of two observing seasons.

The panels in Figure 3 show the distribution of host magnitude and distance for simulated detections. RoSETZ-Lite would be capable of detecting small planets around low-mass stars as faint as F146 $\sim$ 24 (corresponding to distances $\lesssim$ 3 kpc), whilst RoSETZ-Max could go as faint as F146 $\sim$ 25 ($\lesssim$ 4 kpc). These distances are large, though they are not far enough to access planetary systems residing in the Galactic bulge. The RoSETZ rocky HZ planet sample will be confined to systems lying within the Galactic disk. This is, in some ways, advantageous in understanding and characterizing blend systematics that will affect such planets within the GBTDS sample, a point that is discussed further in Section 5.2.

---

[8] https://phl.upr.edu/projects/habitable-exoplanets-catalog/, accessed in May/June 2023.

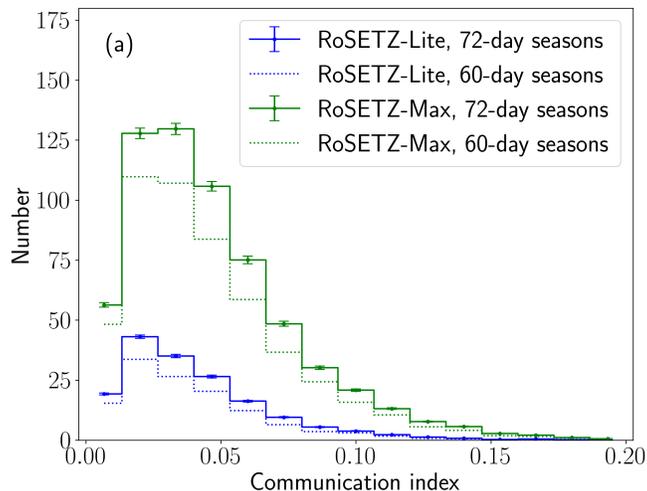

**Figure 5.** The "Communication Index" ($CI$, defined by Equation 3) of detected transit hosts. An index value of unity indicates a host whose angular motion equals the half-width of the ETZ over a timescale equal to the two-way communication time between Earth and the host. Hosts of rocky HZ planets discoverable within the RoSETZ field are predicted to have $CI \ll 1$. This means most remain within the ETZ (and also within the RoSETZ survey area) on a timescale that is much longer than their two-way communication timescale. This is a favourable condition for targeted SETI searches based upon a mutual detectability strategy. Observers on a planet within the RoSETZ survey area, at a time when they saw Earth transit the Sun, and who were incentivized to transmit a signal to us that arrives at Earth now, are likely still to be located within the RoSETZ survey area.

Figure 4 shows the distributions of transit duration and of the host spectral type. The relatively short durations for most detections reflect that we are targeting HZ planets with, mostly, short orbital periods around low-luminosity stars. However, the dense 15 min cadence temporal sampling means that we can expect these transits to be generally very well time resolved. We also see that the sample is dominated by M-dwarf hosts, with a modest number of K-type hosts.

### 4.1 RoSETZ and SETI

As mentioned in Section 2, the RoSETZ exoplanet sample will be the first optimized for future targeted SETI follow-up along principles of Mutual Detectability. In this scenario, civilizations on one of these planets may observe Earth transit the Sun and infer that it is a potentially suitable host for life. They may even be able to observe biomarker signatures or technosignatures in Earth's atmosphere, using techniques that we ourselves use to study exoplanet atmospheres. This, together with their realization that a technological civilization on Earth can glean similar information about them, may be enough incentive for them to send a transmission towards us.

One potential consideration is that, whilst the direction of line of nodes between the Ecliptic and Galactic planes provides many target stars within the ETZ, it is not a stationary point, and neither are stars stationary within it. Heller & Pudritz (2016) argue that, whilst the width of the ETZ varies along the Ecliptic due to the eccentricity of Earth's orbit, and that this variation changes over time as Earth's eccentricity varies, these timescales are long and their effects are relatively small. Planetary precession, due to changes in the orientation of the Ecliptic with respect to the solar system invariable plane, are 100 times smaller than lunisolar precession effects and therefore do not cause significant motion of the ETZ along the Galactic plane.

The importance of stellar proper motions within the RoSETZ field can be considered by comparing the timescale over which a star can travel half of the angular width of the ETZ against the two-way light travel time between Earth and the star. Specifically, we define the Communication Index ($CI$) as:

$$CI = \frac{c\Delta_{\text{ETZ}}}{4|\mu_*|D_*}, \tag{3}$$

where $\Delta_{\text{ETZ}} = 0°\!.53$ is the angular width of the ETZ, $\mu_*$ is the stellar proper motion and $D_*$ is the distance to the star. By construction, stars that traverse an angular distance equal to $\Delta_{\text{ETZ}}/2$ on a timescale equal to the two-way communication time from Earth to the star have $CI = 1$. From Mutual Detectability considerations, it is desirable to have $CI \ll 1$ so that a communication signal we receive today, that was directed at us in response to an historical observation of Earth's transit, originates from a host star that is still likely to be located within the ETZ. Since a single RoSETZ field can span the entire ETZ (c.f. Figure 1), $CI \ll 1$ is also a sufficient condition to ensure that the signal origin likely remains within the RoSETZ field area.

The distribution of $CI$ values for simulated RoSETZ transit detections is shown in Figure 5. Reassuringly, the distribution confirms that all detectable transit systems satisfy this condition. As such, the sample of exoplanets that from RoSETZ is optimized for a Mutual Detectability strategy. The fact that $CI \ll 1$ for the RoSETZ field is testimony to the importance of wide-field surveillance for this work.

## 5 RoSETZ and the GBTDS

The RoSETZ field location lies around 5 degrees away from the nominal GBTDS region. Whilst the GBTDS field placement is not yet finalized, the requirements of the exoplanet microlensing survey will mean that allowable variations in position will be small (Penny et al., 2019).

The encouragement for this White Paper stems from an understanding that the Roman slew and settle times may turn out to be rather less than originally anticipated, allowing the inclusion of additional fields without compromising the science requirements of the Roman GBTDS. It is in this spirit that we propose the RoSETZ-Max design.

Alternatively, the RoSETZ-Lite design may present an attractive option to consider, even if the Roman slew and settle times are not faster than anticipated. In this case, we believe the very small reduction in overall exoplanet microlensing yield, which would result from RoSETZ-Lite taking some time away from the GBTDS, would be compensated by RoSETZ enabling self-calibration of Galactic model systematics that would otherwise hamper exoplanet demographic studies.

We therefore believe that RoSETZ, in some form, provides a net benefit to the overall quality of the GBTDS exoplanet science, in addition to the new science that it enables. We elaborate on this point in what follows.

### 5.1 Calibrating Galactic model systematics in the GBTDS Microlensing Sample

In Figure 6 we show maps of the stellar microlensing rate and average event duration predicted by the Manchester-Besançon Microlensing Simulator (MaB$\mu$LS[9]). This is the most thoroughly tested public model of stellar microlensing that has been shown by Specht et al. (2020) to provide very good agreement with the distribution and properties of the 8,000-event optical microlensing sample published by OGLE (Mróz et al., 2019), which is the largest completeness-corrected microlensing sample published to date. The maps apply to background sources with $K < 23$, corresponding to the majority of background sources to planetary microlens systems that are expected to be detected in the Roman WFI F146 band (Penny et al., 2019). The location of the RoSETZ and nominal GBTDS fields are indicated by white rectangles.

We find that the microlensing event rate within the RoSETZ field, for events with durations below 60 days, would be around 75% that of the lowest yield GBTDS field (at $l = 1°\!.41, b = -1°\!.64$). We find this field contributes ∼12% of the expected total stellar microlensing rate across the seven GBTDS pointings. Normalizing to a GBTDS exoplanet yield of 1,400 planets from 27,000 stellar microlensing events (Penny et al., 2019), we expect RoSETZ-Lite to yield ∼40 exoplanet signals from ∼830 stellar microlensing events. The yields are approximately triple these numbers for RoSETZ-Max. This scaling argument may well be conservative, since blending effects will be less for the RoSETZ field.

---

[9] http://www.mabuls.net/

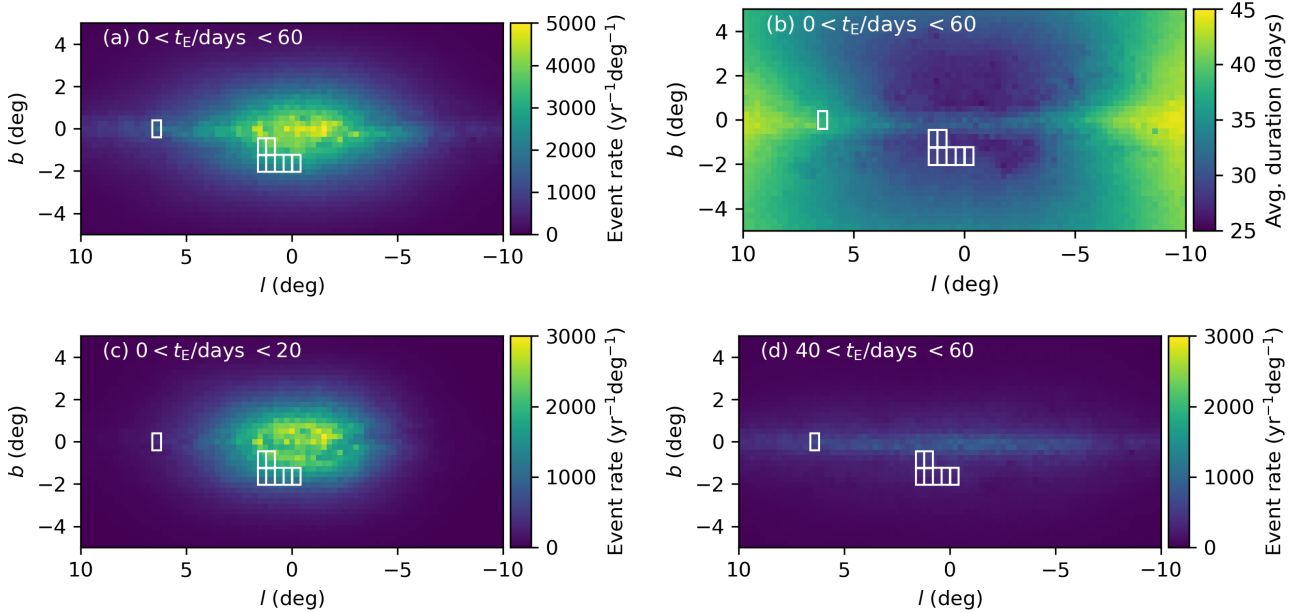

**Figure 6.** The sensitivity of the RoSETZ and GBTDS fields to microlens population. Panel (a) shows the MaBµLS-2[9] simulated $K$-band (2 µm) stellar microlesing rate for events of duration less than 60 days, involving sources brighter than $K = 23$ and having a signal-to-noise ratio greater than 25 at peak (Specht et al., 2020). The nominal seven GBTDS field locations are shown by the cluster of white rectangles close to the Galactic Centre (Penny et al., 2019). The RoSETZ field is shown by the isolated rectangle at $l = 6°.4$, $b = 0°$, and has been given the same vertical orientation for comparison. The total event rate in the RoSETZ field to all background sources (including those fainter than the $K = 23$ limit of these maps) is expected to be $\sim 2,100$ events/yr, compared to an average of $\sim 3,300$ events/yr per pointing within the nominal GBTDS region. Panel (b) shows the average duration of these events is expected to be significantly longer in the RoSETZ field than in the GBTDS region. Panel (c) shows the rate for the subset of events with durations below 20 days, and (d) the rate for events with durations between 40 and 60 days. It is clear from these panels that the longer duration events seen in the RoSETZ field are dominated by lenses in the disk, whilst it is lenses in the bulge that dominate the shorter events seen in the GBTDS. Comparison of the two event samples would provide strong constraints on the Galactic bar geometry. RoSETZ can therefore help enable self-calibration of acceptable Galactic model priors that will need to be used for a substantial fraction of the GBTDS exoplanet sample where a direct mass measurement is unobtainable.

It is worth considering the potential impact of RoSETZ-Lite in the event that the RoSETZ survey could not be accommodated without a commensurate loss of observing time to the GBTDS fields. If two of the six GBTDS observing seasons were devoted to observing the ETZ, at the expense of a two-season reduction of survey time for the lowest-yielding GBTDS field, we would expect the overall (GBTDS+RoSETZ) event yield to reduce by only ∼1% compared to the nominal GBTDS yield.

By contrast, there are potentially significant gains to be had in terms of systematic uncertainties in exoplanet demographic modelling. The current expectation is that at least half of the GBTDS exoplanet microlensing sample will be able to yield direct planet mass measurements due to combinations of microlens parallax, proper motion and lens host flux measurements that allow the microlens mass-distance-velocity degeneracy to be completely broken. This means that up to half of the exoplanet sample will still have at least a partial degeneracy in their mass measurement. Such events are still important, collectively, for exoplanet demography studies, as statistical information on planet masses can be gained through a Bayesian approach using distance and velocity priors from Galactic models. However, the results clearly rest on the reliability of the model prior.

Having a well separated line of sight will provide an important self-calibration of acceptable Galactic model priors. Figure 6 shows that the event timescale is a sensitive diagnostic of the underlying Galactic population from which the lens system originates. Short duration events (below 20 days) are strongly biased towards source and lens stars originating in the bulge, whilst longer duration events (40-60 days) are far more likely to reside in the Galactic disk. This is evidenced by the effective disappearance in Figure 6 of the disk in panel (c) and bulge in panel (d). For the Galactic model used in MaB$\mu$LS, the RoSETZ location essentially only samples the disk population, whilst the nominal GBTDS fields sample events originating from both disk and bulge populations. However, variations in Galactic model parameters (e.g. relative bulge/disk density normalization, bulge/disk luminosity and mass functions, kinematics, orientation of the Galactic bar) will give different results for the relative mix of disk and bulge events towards the RoSETZ and GBTDS fields. These differences will cause observable variations in the relative rate and event timescale distributions, which can be used to constrain allowable Galactic model properties.

Clearly, beyond the immediate science objective of controlling systematics in the interpretation of exoplanet demographics, the Galactic model constraints will enable the stellar phase space of the inner Galaxy to be probed in a more powerful way than can be achieved from the nominal GBTDS region alone. This is a region that, due to stellar crowding, has somewhat limited phase space coverage by the ESA Gaia mission. Stellar microlensing yields from the RoSETZ field can therefore play a pivotal role for the study of inner Galactic structure.

**5.2 Calibrating blending systematics in the GBTDS Transit Sample**

The GBTDS is expected to increase the number of known transiting planets by more than an order of magnitude (Wilson et al., 2023). The transit sample detected by Roman will overlap in distance with the microlensing exoplanet sample, allowing the demographics of hot and cold exoplanet populations to be evaluated among similar hosts. Additionally, Roman will facilitate studies of hot exoplanet demography versus Galactic location via comparisons of the "far-field" Roman sample with the "near-field" Kepler sample.

One potentially important factor for transiting planets in the GBTDS region is blended flux from unresolved background stars lying within the PSF of the host star. Despite the high spatial resolution of Roman, the severe stellar crowding in the bulge means that the PSF of host stars will usually be contaminated by extraneous flux from one or more background stars. This would result in the measured transit depth being an underestimate of the true depth, and therefore the planet radius also being underestimated. The slope of the planet mass-radius relation means that, for rocky planets, a 30% error in the transit depth, whilst translating to just a 15% error in planet radius, implies a greater than 50% error in the inferred planet mass.

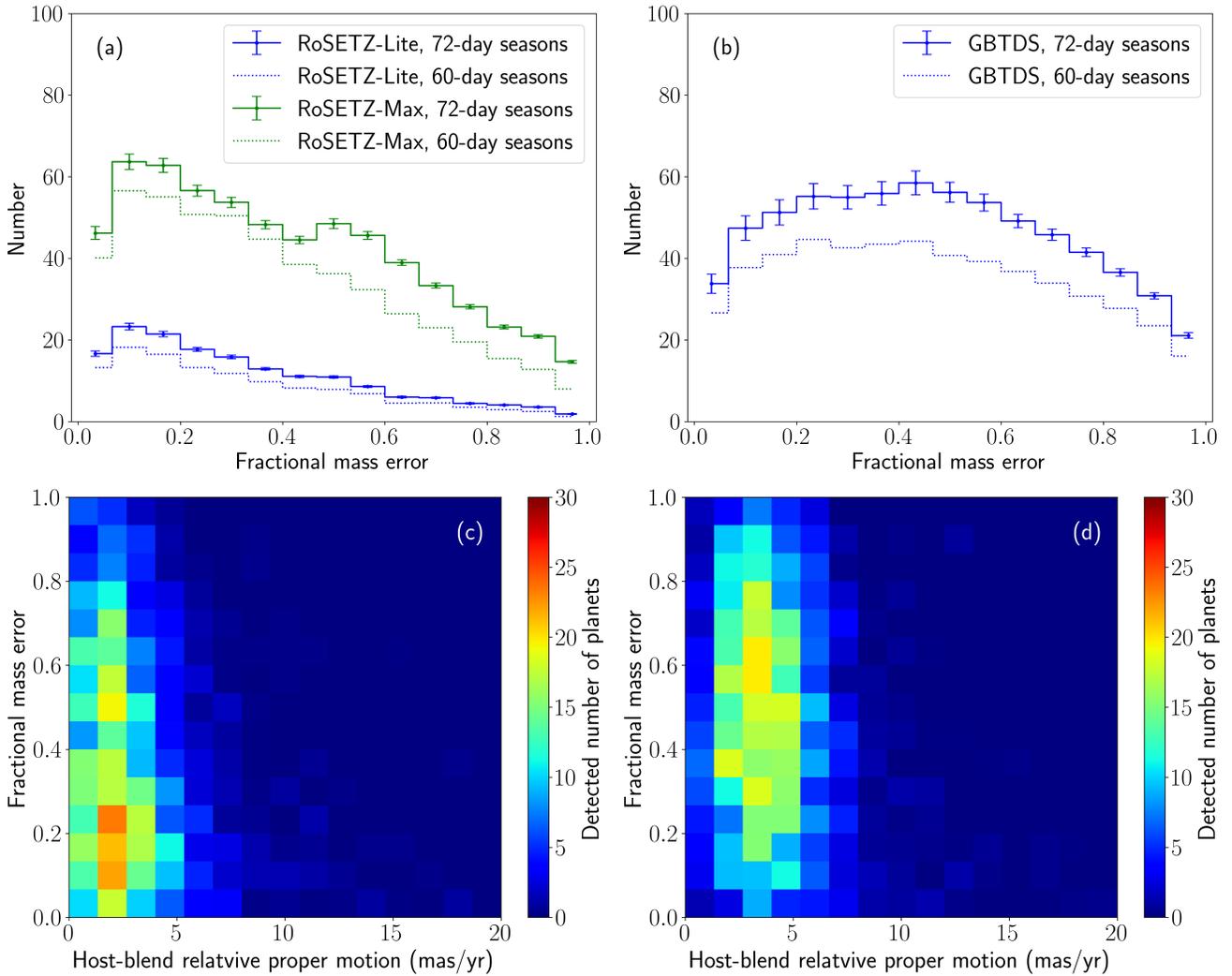

**Figure 7.** Top row: the fractional error on the inferred planet mass [(true-inferred)/true] due to blending for (a) the RoSETZ field, and (b) a representative GBTDS location centred at $l = 0°.5$, $b = -1°.5$. The inferred planet mass is computed from the planet mass-radius relation in Table 1, where the inferred radius is computed from the transit depth, assuming perfect knowledge of the host radius. Bottom row: mass fraction error versus host proper motion relative to the centre-of-light of its blend flux. Over a 4.5-year baseline, hosts with relative proper motion above 2.4 mas will separate from their blends by more than one-tenth of a WFI pixel, which is expected to be comfortably above the achievable centroiding precision for hosts of detectable transits. Panel (c) is for a 72-day season RoSETZ-Max survey of the ETZ, whilst (d) is the equivalent for the representative GBTDS region. In general, the ETZ direction benefits from a lower fractional mass error, whilst the GBTDS fields benefit from faster separation between the host and blend light. The difference in fractional mass error arises directly from differences in blending effects towards the two directions. This difference is Galactic model dependent and provides a basis for Bayesian methods that can statistically correct the planet mass distributions of both samples, even for subsets that cannot be deblended using host–blend proper motion.

The RoSETZ field may offer an important self-calibration of Roman transit measurements in the face of significant blending systematics within the crowded GBTDS area. This can be seen in Figure 7 where the fractional error in the inferred planet mass is shown to be typically rather higher for the representative GBTDS location than for the RoSETZ field. However, the five-year baseline of the GBTDS observations should allow the detection of relative proper motion between the exoplanet host and the centre of light of the blend stars. This is shown plotted against mass error in panels (c) and (d) of Figure 7 for the RoSETZ and GBTDS locations, respectively. Here we see that GBTDS host stars separate from the blend stars at a faster rate than RoSETZ hosts, allowing blending solutions to be achieved for many, though not all, transits. The difference in blending along the two directions is due to the different mix of detectable stars, and so is sensitive to the underlying Galactic model.

As Figure 3 shows, planets detected around low-mass hosts are located within 4 kpc of us and so, for both RoSETZ and GBTDS sight lines, such planetary systems reside in the Galactic disk rather than the bulge. At a distance of 4 kpc, an angular separation of 5° between the RoSETZ and GBTDS fields corresponds to a physical separation of just 0.35 kpc, which is only $\sim 12\%$ of the disk scale length. Under the reasonable assumption that planet demographics should therefore be statistically similar along both locations, the observed difference in the distributions of measured transit depths can be attributed to differences in blending alone. Galactic models that can correctly account for this difference can also be used statistically to "deblend" both samples in order to recover the underlying distribution of planet radii (and masses) from the observed transit depth distributions. In this way, the use of two separated, but close, sight lines can significantly improve the accuracy of the demographics of planets around low-luminosity hosts. Such an approach will be very important for small planet statistics, as these will largely come from small (low luminosity) host stars.

## 6 Other Roman Pillars of Science Enabled by RoSETZ

The location of the RoSETZ field will enable a number of other science opportunities. We briefly discuss a few of them here.

### 6.1 Deep 3D Galactic Extinction Mapping

With large near-IR Galaxy surveys, it has become possible to construct 3D maps of interstellar extinction (Marshall et al., 2006). The methodology rests on the strong correlation between observed $J-K$ colour and distance for subsets of K/M giants and RGB/AGB stars. Using a prior Galactic model a 3D extinction model can be constructed via an iterative procedure until the model and observed $J-K$ distributions match. This was first done for the Besançon Galactic model calibrated against 2MASS data (Marshall et al., 2006). The resulting dust model is correct only to the extent that the input Galactic model is correct. However, Marshall et al. (2006) were able to show that its 2D projection successfully recovered features seen in 2D dust maps.

The Roman F129 and F213 are close proxies to $J$ and $K$ and therefore could be used to construct 3D dust maps using the same method. Occasional sampling in these filters would also provide useful colour information for transit false-positive identification. With almost 6,000 epochs collected over a 60-day observing season, even if only $\sim 5\%$ were used for F129 and F213 observations, the depth and effective pixel resolution of the dithered stack would enable much deeper and higher angular resolution maps of the dust distribution within the RoSETZ field than hitherto possible.

### 6.2 Searches for Centaurs and Trans Neptunian Objects

The dynamics and characteristics of populations of small solar system objects hold vital clues to the formation of the solar system. But such bodies are also very faint and so intrinsically difficult to detect. Mid and near-IR searches have been used to look for Trans Neptunian Objects (TNOs) with orbits above 30 au ([Fernández-Valenzuela et al., 2021](#)), and Centaurs that are located between the orbits of Jupiter and Neptune ([Delsanti et al., 2004](#); [Hainaut et al., 2012](#)).

RoSETZ may present a good opportunity to look for such objects through astrometry performed on very deep drizzled image stacks. With around 6000 exposures over a 60-day period, each with a 5$\sigma$ point source limit of $F146 \simeq 25.7$, $F213 \simeq 23.7$ Roman can provide extremely high resolution deep views. A 1-hour stack of 65 exposures can reach $F213 \simeq 26.1$, which compares to magnitudes of observed Centaur members of $K \sim 18-20$ (e.g. [Delsanti et al., 2004](#)). With $\sim 1,500$ hours observations per season there is a fantastic opportunity to compile a large sensitive database of magnitudes and orbit solutions for Centaur and TNO candidates, some of which could be followed-up with JWST mid-IR observations for compositional studies ([Fernández-Valenzuela et al., 2021](#)).

# 7 Summary

We propose the *Roman Survey of the Earth Transit Zone* (RoSETZ), a transit search for rocky planets within the habitable zones (HZs) of stars located within the Earth Transit Zone (ETZ), where observers on those planets could observe Earth transiting the Sun. RoSETZ would augment the Roman Galactic Bulge Time Domain Survey (GBTDS) as an additional field located $\sim 5$ degrees away from other GBTDS fields. RoSETZ can find hundreds of Earth-sized HZ planets around K- and M-type hosts, with around 120 forecast for a RoSETZ-Lite two-season design, and 630 for a RoSETZ-Max six-season survey running as part of the GBTDS. These yields are 5–20 times the number currently known.

Such a large sample will transform our knowledge of the occurrence of Earth-sized HZ planets. It would also be the first catalogue of exoplanets selected in a manner optimized for follow-up searches for extra-terrestrial intelligence (SETI). Most if not all the Earth-sized HZ planets would be located within the ETZ and would be good targets for future SETI surveys according to the recent Mutual Detectability strategy for SETI searches. Given the huge public interest in the possibility of life beyond Earth, there are many outreach opportunities that could come from the planet sample observed by RoSETZ.

If it can be accommodated alongside the existing GBTDS design, we favour the RoSETZ-Max design that is observed for the duration of the GBTDS. If not, the slimmed-down RoSETZ-Lite design would not significantly impact overall exoplanet yields, even if time allocated to it had to come from time allocations to other fields. We argue that the angular separation of RoSETZ from other GBTDS fields permits self-calibration of systematic uncertainties that would otherwise hamper exoplanet demographic modelling of both microlensing and transit datasets. Lastly, several other possible science areas could be facilitated by the deep, high-resolution RoSETZ dataset.

## Acknowledgements

R. H. acknowledges support from the German Aerospace Agency (Deutsches Zentrum für Luft- und Raumfahrt) under PLATO Data Center grant 50OO1501. S.S. acknowledges the SETI Forward Award from the SETI Institute.

## References

Borucki W. J., et al., 2010, Science, 327, 977
Bryson S., et al., 2021, AJ, 161, 36
Burn R., Schlecker M., Mordasini C., Emsenhuber A., Alibert Y., Henning T., Klahr H., Benz W., 2021, A&A, 656, A72
Delsanti A., Hainaut O., Jourdeuil E., Meech K. J., Boehnhardt H., Barrera L., 2004, A&A, 417, 1145
Eker Z., et al., 2018, MNRAS, 479, 5491
Fernández-Valenzuela E., et al., 2021, Planetary Society Journal, 2, 10
Garrett M. A., Siemion A. P. V., 2023, MNRAS, 519, 4581

Hainaut O. R., Boehnhardt H., Protopapa S., 2012, A&A, 546, A115

Haqq-Misra J., et al., 2022, Acta Astronautica, 198, 194

Hardegree-Ullman K. K., Cushing M. C., Muirhead P. S., Christiansen J. L., 2019, AJ, 158, 75

Heller R., Pudritz R. E., 2016, Astrobiology, 16, 259

Heller R., et al., 2014, Astrobiology, 14, 798

Kerins E., 2021, AJ, 161, 39

Kerins E., Robin A. C., Marshall D. J., 2009, MNRAS, 396, 1202

Kipping D. M., Teachey A., 2016, MNRAS, 459, 1233

Kopparapu R. K., Ramirez R. M., SchottelKotte J., Kasting J. F., Domagal-Goldman S., Eymet V., 2014, ApJ, 787, L29

Lacki B. C., et al., 2021, ApJS, 257, 42

Marshall D. J., Robin A. C., Reylé C., Schultheis M., Picaud S., 2006, A&A, 453, 635

McDonald I., et al., 2014, MNRAS, 445, 4137

Mróz P., et al., 2019, ApJS, 244, 29

Pascucci I., Mulders G. D., Gould A., Fernandes R., 2018, ApJ, 856, L28

Penny M. T., et al., 2013, MNRAS, 434, 2

Penny M. T., Gaudi B. S., Kerins E., Rattenbury N. J., Mao S., Robin A. C., Calchi Novati S., 2019, ApJS, 241, 3

Pinamonti M., et al., 2022, A&A, 664, A65

Price D. C., et al., 2020, AJ, 159, 86

Rauer H., et al., 2014, Experimental Astronomy, 38, 249

Ricker G. R., et al., 2015, Journal of Astronomical Telescopes, Instruments, and Systems, 1, 014003

Robin A. C., Reylé C., Derrière S., Picaud S., 2003, A&A, 409, 523

Robin A. C., Marshall D. J., Schultheis M., Reylé C., 2012, A&A, 538, A106

Sabotta S., et al., 2021, A&A, 653, A114

Sheikh S. Z., Siemion A., Enriquez J. E., Price D. C., Isaacson H., Lebofsky M., Gajjar V., Kalas P., 2020, AJ, 160, 29

Specht D., Kerins E., Awiphan S., Robin A. C., 2020, MNRAS, 498, 2196

Spergel D., et al., 2015, arXiv e-prints, p. arXiv:1503.03757

Wilson R. F., et al., 2023, arXiv e-prints, p. arXiv:2305.16204

Wright J. T., 2018, in Deeg H. J., Belmonte J. A., eds, , Handbook of Exoplanets. p. 186, doi:10.1007/978-3-319-55333-7_186